\documentstyle [twoside,12pt]{article}
\begin{document}
\setlength{\baselineskip}{.7cm}
\sloppy
\begin{center}
{\large \bf Squeezed Neutrino Oscillations}
\end{center}
\begin{center}
{\large \bf in Quantum Field Theory}
\end{center}
\centerline{ E.Alfinito$^{\dag}$
\footnote{alfinito@le.infn.it}, M. Blasone
\footnote{blasone@vaxsa.csied.unisa.it},
A.Iorio\footnote{iorio@vaxsa.csied.unisa.it}
and G. Vitiello\footnote{vitiello@vaxsa.csied.unisa.it}}

\vspace{.2in}
\centerline{Dipartimento di Fisica dell'Universit\`a}
\centerline{and INFN, Gruppo Collegato, Salerno}
\centerline{I-84100 Salerno, Italy}

\vspace{.1in}
\centerline{${}^{\dag}$Dipartimento di Fisica dell'Universit\`a}
\centerline{Via Arnesano}
\centerline{I-73100 Lecce, Italy}

\medskip
\medskip
\medskip
\medskip
\begin{flushright}
SADF1-1995****
\end{flushright}
\sloppy

\begin{abstract}
By resorting to recent results on fermion mixing
which show that the Fock space of
definite flavor states is unitarily inequivalent to the Fock space
of definite mass states, we discuss the phenomenological implications on
the neutrino oscillation formula.
For finite momentum the oscillation amplitude is depressed, or "squeezed",
by a momentum dependent factor. In the
relativistic limit the conventional oscillation formula is recovered.
\end{abstract}

P.A.C.S. ****** 11.10.-z , 11.30.Jw , 14.60.Gh

\thispagestyle{empty}

\newpage
In a recent paper \cite{1} the mixing transformations of fermion fields
have been
studied in the framework of the Lehmann-Symanzik-Zimmermann (LSZ) formalism
of quantum field theory (QFT) \cite{2} and particular attention has been
devoted
to the mixing transformations of massive Dirac neutrino fields in view of
their relevance to neutrino physics and related questions of great interest
to cosmology as well as to solar physics and modelling \cite{3}.

The analysis presented in ref. 1 shows that much care is needed in the
identification of the proper vacuum state for the mixed fields since
the Fock space for the original (free) fields turns
out to be unitarily inequivalent to the Fock space for the
mixed fields in the infinite volume
limit.

Difficulties in defining the appropriate creation and annihilation
operators for mixed neutrino fields such as the ones which are used in the
standard treatment of neutrino oscillations were already pointed out in
\cite{4} . There it was shown that it is in fact impossible to construct
operators for weak states which obey canonical anticommutation relations.
"Approximate" operators and the corresponding
"approximate Fock space" were constructed which were shown to exist
only in the relativistic
limit and for almost degenerate
mass eigenvalues. In the light of the results presented  in
ref. 1 it now appears that the difficulties pointed out in ref. 4
may find their origin in the unitary
inequivalence between the mixed fields and the massive
(free) fields Fock spaces. Creation and annihilation operators
for the mixed fields which satisfy canonical
anticommutation relations are explicitly constructed in ref. 1
and the vacuum state for a well definite Fock space is found to be an SU(2)
generalized coherent state with neutrino-antineutrino condensate structure.

The results of ref. 1, when applied to neutrino mixing,
lead to non-trivial consequences in the neutrino oscillation formula, as
we will explain below. The question then arises if and to which extent
such results may change the experimental expectations and may be eventually
tested. The purpose of this paper is indeed to discuss such a question and
to estimate the corrections to the conventional neutrino oscillation
formula.

For the reader convenience , let us briefly summarize the results of ref. 1.
We will omit all the mathematical analysis and derivations which are there
reported in detail.

We consider the Pontecorvo mixing relations \cite{5} (for simplicity we
confine
ourselves to two flavors; for the case of three flavors see ref. 1):
$$\nu_{e}(x) = \nu_{1}(x) \; \cos\theta   + \nu_{2}(x) \; \sin\theta  $$
$$\nu_{\mu}(x) =- \nu_{1}(x) \; \sin\theta   + \nu_{2}(x)\;
\cos\theta\;,  \eqno(1)$$
where $\nu_{e}(x)$ and $\nu_{\mu}(x)$ are the (Dirac) neutrino fields
with definite flavors.
$\nu_{1}(x)$ and $\nu_{2}(x)$ are the (free) neutrino
fields with definite masses $m_{1}$ and $m_{2}$, respectively.
Here we do not need to distinguish between left-handed and right-handed
components.
The fields $\nu_{1}(x)$ and $\nu_{2}(x)$ are written as
$$\nu_{i}(x) = \frac{1}{\sqrt{V}} \sum_{\bar{k},r}[u^{r}_{\bar{k},i}
\alpha ^{r}_{\bar{k},i}\:e^{i \bar{k}\cdot \bar{x}}+ v^{r}_{\bar{k},i}
\beta ^{r\dag }_{\bar{k},i}\: e^{-i \bar{k}\cdot \bar{x}}], \; ~ i=1,2 \;.
\eqno(2) $$
$\alpha ^{r}_{k,i}$ and $ \beta ^{r }_{k,i}$, $  i=1,2 \;,
\;r=1,2$ are the annihilator operators for the vacuum state
$|0\rangle_{1,2}$:
$\alpha ^{r}_{k,i}|0\rangle_{12}= \beta ^{r }_{k,i}|0\rangle_{12}=0$.
Here and in the following, as far as no misunderstanding arises, we omit time
dependence. The anticommutation relations are:
$$\{\nu^{\alpha}_{i}(x), \nu^{\beta\dag }_{j}(y)\}_{t=t'} = \delta^{3}(x-y)
\delta _{\alpha\beta} \delta_{ij} \;, \;\;\;\;\; \alpha,\beta=1,..,4 \;,
\eqno(3)$$
and
$$\{\alpha ^{r}_{k,i}, \alpha ^{s\dag }_{q,j}\} = \delta
_{kq}\delta _{rs}\delta _{ij}   ;\qquad \{\beta ^{r}_{k,i}, \beta ^{s\dag
}_{q,j}\} = \delta _{kq} \delta _{rs}\delta _{ij},\;\;\;\;
i,j=1,2\;. \eqno(4)  $$
All other anticommutators are zero. The orthonormality and
completeness relations are the usual ones.

Eqs.(1) relate the hamiltonians $H_{1,2}$ (we consider only the mass
terms) and $H_{e,\mu}$ \cite{5}:
$$H_{1,2}=m_{1}\;\bar {\nu}_{1} \nu_{1} + m_{2}\;\bar {\nu}_{2} \nu_{2}
\eqno(5)$$
$$H_{e,\mu}=m_{ee}\; \bar {\nu}_{e} \nu_{e} +
m_{\mu\mu}\;\bar {\nu}_{\mu} \nu_{\mu}+
m_{e\mu}\left(\bar {\nu}_{e} \nu_{\mu} + \bar {\nu}_{\mu} \nu_{e}\right)
\eqno(6)$$
where $m_{ee}=m_{1}\cos^{2}\theta + m_{2} \sin^{2} \theta$,
$m_{\mu\mu}=m_{1}\sin^{2}\theta + m_{2} \cos^{2} \theta$
and $m_{e\mu}=(m_{2}-m_{1})\sin\theta \cos \theta$.

It is useful to mention at this point that in the LSZ formalism of QFT
\cite{2}
asymptotic in- (or out-) fields, also called free or physical fields, in
terms of which observables are expressed, are obtained by the weak limit of
the Heisenberg or interacting fields for $t \rightarrow - (or +)
\infty$ . The basic dynamics, namely the system Lagrangian and the
resulting field equations, is given in terms of the Heisenberg fields and
therefore the meaning of the weak limit is to provide a realization of the
basic dynamics in terms of the asymptotic fields. Such a realization, i.e.
the weak limit, is however not unique since infinitely many representations
of the canonical (anti-)commutation relations exist in QFT \cite{2,6,7}. Well
known examples of such a situation are the theories where spontaneous
breakdown of symmetry is possible. There, the same set of Heisenberg fields
and the same basic dynamics can be realized by asymptotic limit in the
normal (symmetric) phase as well as in the broken symmetry phase.
Therefore, since unitarily inequivalent representations describe physically
different phases, in order to avoid ambiguities, it is of crucial
importance to investigate with much care the mapping among Heisenberg
fields and free fields (these mappings are usually called dynamical
mappings or Haag expansions \cite{6,7}).

The above remarks apply to QFT, where systems with infinite
number of degrees of freedom are considered.
In quantum mechanics, namely for finite volume
systems, the von Neumann theorem ensures that the
representations of the canonical commutation relations
are each other unitary equivalent and no problem arises
with uniqueness of the asymptotic limit. However, the von Neumann
theorem does not hold in QFT and much attention is required when
considering any mapping among interacting and free fields \cite{6,7}.

For these reasons, intrinsic to the QFT structure,
mixing relations such as the relations (1) deserve a
careful analysis.

It was in fact the purpose of ref. 1 to investigate  the
structure of the Fock spaces ${\cal H}_{1,2}$ and ${\cal H}_{e,\mu}$
relative to
$\nu_{1}(x)$, $\nu_{2}(x)$ and  $\nu_{e}(x)$,  $\nu_{\mu}(x)$, respectively.
In particular, there it was indeed shown that the massive fields space
${\cal H}_{1,2}$ and the flavor fields space ${\cal H}_{e,\mu}$
become orthogonal (i.e. unitarily inequivalent) in the infinite volume
limit: $\lim_{V \rightarrow \infty}\; _{1,2}\langle0|0\rangle_{e,\mu} =
 0 $~, where $|0\rangle_{e,\mu}$ denotes the vacuum for the flavor field
operators.  This is an exact result in
QFT and is a novel feature with respect to the conventional treatment of
neutrino mixing.

The unitary inequivalence in the infinite volume limit of the mass
and the flavor representations shows the absolutely non-trivial nature of
the mixing transformations (1).
In fact, one can show that the mixing
transformations induce a physically non-trivial structure in the flavor
vacuum state which indeed turns out to be an $SU(2)$ generalized coherent
state \cite{8}
exhibiting neutrino-antineutrino pair condensation \cite{1}.

We thus realize the
limit of validity of the approximation usually adopted when the vacuum
state of the representation for definite mass operators is identified
with the vacuum state for the flavor operators.
We point out that even at finite volume the vacua identification is actually
an approximation since the flavor vacuum is an $SU(2)$ generalized coherent
state. In such an approximation, the coherent state structure
with pair condensation is in fact missed.

In conclusion, only in a theoretically
rude approximation one may assume that massive neutrino fields and flavor
neutrino
fields share the same vacuum state and the same Fock space representation.
The problem is, however, to see if the proper theoretical treatment leads
to any interesting and testable effect out of reach in the
heuristic conventional approximation. Our following discussion is aimed
to such a task.

Without loss of generality, one can choose \cite{1} the reference
frame such that
$k=(0,0,|k|)$. The mixing transformations (1) then lead to the mappings
in terms of creation and annihilation operators \cite{1}:
$$\alpha^{r}_{k,e}=\cos\theta\;\alpha^{r}_{k,1}\;+\;\sin\theta\;\left(
U_{k}^{*}\; \alpha^{r}_{k,2}\;+\;\epsilon^{r}\;
V_{k}\; \beta^{r\dag}_{-k,2}\right) \eqno(7a)$$
$$\alpha^{r}_{k,\mu}=\cos\theta\;\alpha^{r}_{k,2}\;-\;\sin\theta\;\left(
U_{k}\; \alpha^{r}_{k,1}\;-\;\epsilon^{r}\;
V_{k}\; \beta^{r\dag}_{-k,1}\right)  \eqno(7b)$$
$$\beta^{r}_{-k,e}=\cos\theta\;\beta^{r}_{-k,1}\;+\;\sin\theta\;\left(
U_{k}^{*}\; \beta^{r}_{-k,2}\;-\;\epsilon^{r}\;
V_{k}\; \alpha^{r\dag}_{k,2}\right) \eqno(7c)$$
$$\beta^{r}_{-k,\mu}=\cos\theta\;\beta^{r}_{-k,2}\;-\;\sin\theta\;\left(
U_{k}\; \beta^{r}_{-k,1}\;+\;\epsilon^{r}\;
V_{k}\; \alpha^{r\dag}_{k,1}\right) \eqno(7d)$$
with $\epsilon^{r}=(-1)^{r}$ and
$$V_{k}=|V_{k}|\;e^{i(\omega_{k,2}+\omega_{k,1})t}\;\;\;\;,\;\;\;\;
U_{k}=|U_{k}|\;e^{i(\omega_{k,2}-\omega_{k,1})t} \eqno(8)$$
$$|U_{k}|=\left(\frac{\omega_{k,1}+m_{1}}{2\omega_{k,1}}\right)^{\frac{1}{2}}
\left(\frac{\omega_{k,2}+m_{2}}{2\omega_{k,2}}\right)^{\frac{1}{2}}
\left(1+\frac{k^{2}}{(\omega_{k,1}+m_{1})(\omega_{k,2}+m_{2})}\right)
\eqno(9a)$$
$$|V_{k}|=\left(\frac{\omega_{k,1}+m_{1}}{2\omega_{k,1}}\right)^{\frac{1}{2}}
\left(\frac{\omega_{k,2}+m_{2}}{2\omega_{k,2}}\right)^{\frac{1}{2}}
\left(\frac{k}{(\omega_{k,2}+m_{2})}-\frac{k}{(\omega_{k,1}+m_{1})}\right)
\eqno(9b)$$
$$|U_{k}|^{2}+|V_{k}|^{2}=1 \eqno(10)$$
$$|V_{k}|^{2}=|V(k,m_{1}, m_{2})|^{2}=
\frac{k^{2}\left[(\omega_{k,2}+m_{2})-(\omega_{k,1}+m_{1})\right]^{2}}
{4\;\omega_{k,1}\omega_{k,2}(\omega_{k,1}+m_{1})(\omega_{k,2}+m_{2})}\eqno
(11)$$
where $\omega_{k,i}=\sqrt{k^{2}+m_{i}^{2}}$.

Most part of our discussion in this paper will be focused on the function
$|V_{k}|^{2}$.

We notice that from eqs.(7) the expectation value of the number
operator $N_{\sigma_{l}}^{k,r}$ is obtained as:
$$\;_{1,2}\langle0|\;N_{\sigma_{l}}^{k,r}\;|0\rangle_{1,2}\;=
\;\sin^{2}\theta\;|V_{k}|^{2} \;,\;\;\;\; \sigma=\alpha, \beta \;,\;\;\;\;
l=e,\mu ,\eqno(12)$$
in contrast with the usual approximation case where one puts
$|0\rangle_{e,\mu}=|0\rangle_{1,2}
\equiv|0\rangle$ and it is $\langle0|\;N_{\alpha_{e}}^{k,r}\;
|0\rangle\;=\;\langle0|\;N_{\alpha_{\mu}}^{k,r}\;
|0\rangle\;=0\;$.
Eq.(12) gives the condensation density of the vacuum state
$|0\rangle_{1,2}$
as a function of the mixing angle $\theta$, of the
masses $m_{1}$ and $m_{2}$, and of the momentum modulus $k$.
It has been also observed that
$_{1,2}\langle0|\;N_{\sigma_{l}}^{k,r}\;|0\rangle_{1,2}$
plays the role of zero point contribution when considering
the energy contribution of
${\sigma_{l}}^{k,r}$ particles \cite{1}.

The oscillation formula is finally obtained by using the mixing mappings
(7) \cite{1}:
$$\langle\alpha_{k,e}^{r}(t)|\;N_{\alpha_{e}}^{k,r}\;
|\alpha_{k,e}^{r}(t)\rangle\;=$$
$$=\;1 -\;\sin^{2}\theta\;|V_{k}|^{2}\;
-\;|U_{k}|^{2}\;\sin^{2}2\theta
\;\sin^{2}\left(\frac{\Delta\omega_{k}}{2}t\right)\;. \eqno(13)$$

The fraction of $\alpha_{\mu}^{k,r}$ particles in the same
state is
$$\langle\alpha_{k,e}^{r}(t)|\;N_{\alpha_{\mu}}^{k,r}\;
|\alpha_{k,e}^{r}(t)\rangle\;=$$
$$=\;|U_{k}|^{2}\;\sin^{2}2\theta
\;\sin^{2}\left(\frac{\Delta\omega_{k}}{2}t\right)+
\;\sin^{2}\theta\;|V_{k}|^{2}\;
\left(1\;-\;\sin^{2}\theta\;|V_{k}|^{2}\right)\;.\eqno(14)$$

The occurrence of $|V_{k}|^{2}$ and of $|U_{k}|^{2}$ in (13) and (14)
denote the contribution from the vacuum condensate.
We observe that
$$\langle\alpha_{k,e}^{r}(t)|\;N_{\alpha_{e}}^{k,r}\;
|\alpha_{k,e}^{r}(t)\rangle +
\langle\alpha_{k,e}^{r}(t)|\;N_{\alpha_{\mu}}^{k,r}
\;|\alpha_{k,e}^{r}(t)\rangle=
$$
$$
\langle\alpha_{k,e}^{r}|\;N_{\alpha_{e}}^{k,r}\;
|\alpha_{k,e}^{r}\rangle +
\langle\alpha_{k,e}^{r}|\;N_{\alpha_{\mu}}^{k,r}
\;|\alpha_{k,e}^{r}\rangle\;. \eqno(15)$$
where $|\alpha_{k,e}^{r}\rangle = |\alpha_{k,e}^{r}(t = 0)\rangle$, which
shows the conservation of the number $(N_{\alpha_{e}}^{k,r}\;
+ N_{\alpha_{\mu}}^{k,r})$ . Notice that the expectation value of this
number in the state $|0\rangle_{1,2}$ is not zero due to the condensate
contribution.

Eqs.(13) and (14) are to
be compared with the approximated ones in the conventional treatment:
$$\langle\alpha_{k,e}^{r}(t)|\;N_{\alpha_{e}}^{k,r}\;
|\alpha_{k,e}^{r}(t)\rangle\;=\;
1-\sin^{2}2\theta\;\sin^{2}\left(\frac{\Delta\omega_{k}}{2}t
\right)\;  \eqno(16)$$
and
$$\langle\alpha_{k,e}^{r}(t)|\;N_{\alpha_{\mu}}^{k,r}\;
|\alpha_{k,e}^{r}(t)\rangle\;=\;
\sin^{2}2\theta\;\sin^{2}\left(\frac{\Delta\omega_{k}}{2}t
\right)~, \eqno(17)$$
respectively.

The QFT results (13) and (14) reproduces the conventional ones
(16) and (17)
when $|U_{k}|\rightarrow 1$ (and $|V_{k}|\rightarrow 0$).

In conclusion, eqs.(13) and (14) exhibit the corrections to the
flavor oscillations
coming from the condensate contributions.
The conventional (approximate) results (16) and (17) are
obtained when the condensate
contributions are missing (in the $|V_{k}| \rightarrow 0$ limit).

To discuss the phenomenological implications of
the results (13) and (14) we have to study the function
$|V_{k}|^{2}$.

Let us immediately observe that $|V_{k}|^{2}$  depends on $k$ only
through its modulus and it is always in the interval $[0,\frac{1}{2}[$.
It has a maximum for $k= \sqrt{m_{1}m_{2}}$.
Also, $|V_{k}|^{2} \rightarrow 0$ when $k \rightarrow \infty$.
Moreover, $|V_{k}|^{2}=0$  when $m_{1} = m_{2}$~
(no mixing occurs in Pontecorvo theory in this case).

It is remarkable that the corrections to the oscillations depend on the
modulus $k$ through $|V_{k}|^{2}$ (and $|U_{k}|^{2} =1-|V_{k}|^{2}$).
Since $|V_{k}|^{2} \rightarrow 0$ when $k \rightarrow \infty$,
these corrections disappear in the infinite momentum or relativistic
($k >> \sqrt{m_{1} m_{2}}$)
limit. However,
for finite $k$, the oscillation amplitude is depressed, or "squeezed",
by a factor
$|U_{k}|^{2}$: the squeezing factor ranges
from $1$ to $\frac{1}{2}$ depending on $k$ and on the masses values.
The values of the squeezing factor may therefore
have not
negligible effects in experimental findings and the dependence
of the flavor oscillation amplitude on the momentum could thus be tested.

To better estimate the effects of the momentum dependence it is useful to
rewrite the $|V_{k}|^{2}$ function as
$$|V_{k}|^{2} \equiv |V(p,a)|^{2}=\frac{1}{2}\left(1-\frac{(p^{2}+1)}
{\sqrt{(p^{2}+1)^{2}+a\;p^{2}}}\right)=
\frac{1}{2}\left(1-\frac{1}{\sqrt{1+a\left(\frac{p}{p^{2}+1}\right)^{2}}}
\right)  \eqno(18)$$
with
$$p=\frac{k}{\sqrt{m_{1} m_{2}}}\;\;\;\;\;,\;\;\;\;a=\frac{(\Delta m)^{2}}
{m_{1} m_{2}}\;\;\;,\;\;\;0\leq a < + \infty ~, \eqno(19)$$
where $\Delta m \equiv m_{2}-m_{1}$ (we take $m_{1}\leq m_{2}$).

At $p=1$, $|V(p,a)|^{2}$ reaches its maximum value $|V(1,a)|^{2}$, which goes
asymptotically to 1/2 when $a \rightarrow \infty$.

It is useful to calculate the value of $p$, say $ p_{\epsilon}$, at which
the function $|V(p,a)|^{2}$ becomes a
fraction $\epsilon$ of its maximum value
$V(1,a)$. This can be obtained by solving the equation:
$$|V(p_{\epsilon},a)|^{2}=\epsilon \; |V(1,a)|^{2} ~.\eqno(20)$$
The solution of this equation is
$$p_{\epsilon}=\sqrt{-c+\sqrt{c^{2}-1}}\;\;\;\;,\;\;\;\;
c\equiv\frac{b^{2}(a+2)-2}{2(b^{2}-1)} \;\;\;\;,\;\;\;\;
b\equiv1-\epsilon\left(1-\frac{2}{\sqrt{a+4}}\right)~.\eqno(21)$$

In Tab. 1 are reported the values of  $\sqrt{ m_{1} m_{2} } $ and of $a$
corresponding to some given values of $m_{1}$ and $m_{2}$ chosen below the
current experimental bounds.

In Tab. 2 three sets of values of $|U(p_{\epsilon},a)|^{2}$ and of
$k_{\epsilon}$, for $\epsilon=1\, , \, \frac{1}{2}\, , \, \frac{1}{10}$,
corresponding to the values of $m_{1}$ and $m_{2}$ given in Tab. 1, are
reported (see also Fig. 1).
We used $|U(p_{\epsilon},a)|^{2}=1-\epsilon+\epsilon \;|U(1,a)|^{2}$ and
$k_{\epsilon}=p_{\epsilon}\; \sqrt{m_{1} m_{2}}$.

We see that for neutrinos of not very large momentum sensible squeezing
factors for the oscillation amplitudes may be obtained.
Larger deviations from the usual oscillation formula may thus be expected
in these low momentum ranges. We note that observations of
neutrino oscillations by large passive detectors include neutrino momentum as
low as few hundreds of KeV \cite{3}.

We observe that the functional dependence of the oscillating amplitude on the
momentum is such that, if experimentally tested, may give an indication on the
neutrino masses since the function $|U_{k}|^{2}$
(cf. eqs.(10) and (13)) has a minimum at $k= \sqrt{m_{1}m_{2}}$.

\begin{table}[b]
\caption{}
\vspace{0.4cm}
\begin{center}
\begin{tabular}{|c||c|c|c|c|} \hline\hline
&{\em $m_{1}$(eV)} & {\em $m_{2}$(KeV)} & {\em $\sqrt{m_{1}m_{2}}$(KeV)} &
{\em $a$}  \\ \hline
$A$ & $ 5   $   &  $ 250 $    &  $   1.12    $    &  $ \sim 5 \cdot 10^{4}  $
\\ \hline
$B$ & $ 2.5   $   &  $ 250 $    &  $   0.79    $    &  $ \sim 1 \cdot 10^{5}  $
\\ \hline
$C$ & $ 5   $   &  $ 200 $    &  $   1    $    &  $ \sim 4 \cdot 10^{4}  $
\\ \hline
$D$ & $  1  $   &  $ 100 $    &  $  0.32  $  &  $ \sim 1 \cdot 10^{5}  $  \\
\hline
$E$ & $ 0.5 $   &  $ 50  $    &  $  0.15  $  &  $ \sim 1 \cdot 10^{5}  $  \\
\hline
$F$ & $ 0.5 $   &  $ 1   $    &  $  0.02  $  &  $ \sim 2 \cdot 10^{3}  $  \\
\hline
\hline
\end{tabular}
\end{center}
\end{table}


\begin{table}[t]
\caption{}
\vspace{0.4cm}
\begin{center}
\begin{tabular}{|c||c|c||c|c||c|c|} \hline\hline

& {\em $|U(1,a)|^{2}$}           &{\em $k_{1}$(KeV)}
& {\em $|U(p_{1/2},a)|^{2}$}     & {\em $k_{1/2}$(KeV)}
& {\em $|U(p_{1/10},a)|^{2}$}    & {\em $k_{1/10}$(KeV)} \\ \hline
$A$ & $ \simeq 0.5 $  &  $ 1.12 $    &  $ \simeq 0.75 $  &  $ \simeq 146 $
&  $ \simeq 0.95 $  &  $ \simeq 519 $ \\ \hline
$B$ & $ \simeq 0.5 $  &  $ 0.79 $    &  $ \simeq 0.75 $  &  $ \simeq 145 $
&  $ \simeq 0.95 $  &  $ \simeq 518 $ \\ \hline
$C$ & $ \simeq 0.5 $  &  $ 1 $    &  $ \simeq 0.75 $  &  $ \simeq 117 $
&  $ \simeq 0.95 $  &  $ \simeq 415 $ \\ \hline
$D$ & $ \simeq 0.5 $  &  $ 0.32 $  &  $ \simeq 0.75 $  &  $ \simeq  58 $
&  $ \simeq 0.95 $  &  $ \simeq 206 $ \\ \hline
$E$ & $ \simeq 0.5 $  &  $ 0.16 $  &  $ \simeq 0.75 $  &  $ \simeq  29 $
&  $ \simeq 0.95 $  &  $ \simeq 104 $ \\ \hline
$F$ & $ \simeq 0.5 $  &  $ 0.02 $  &  $ \simeq 0.75 $  &  $ \simeq 0.6 $
&  $ \simeq 0.95 $  &  $ \simeq  2  $ \\ \hline
\hline
\end{tabular}
\end{center}
\end{table}

It is interesting to observe that, although
in a different framework where the neutrino wave
packet spreading and the effect of spatially localized source and detectors
were studied, also in refs. 9 it has been
pointed out
that non relativistic neutrinos with different masses are expected to
give rise to drastically depressed oscillation amplitudes, the usual
oscillation formula being recovered in the relativistic limit.
As a conclusion, probing
the non relativistic momentum domain seems therefore promising
in order to obtain new insights in neutrino physics.

Since, as we have shown,
the correction factor is related to the vacuum condensate,
we see that the vacuum acts as a "momentum (or spectrum) analyzer" for the
oscillating neutrinos: neutrinos with $k\gg\sqrt{m_{1}m_{2}}$
have oscillation amplitude larger
than neutrinos with $k\simeq\sqrt{m_{1}m_{2}}$,
due to the vacuum structure. Such a
vacuum spectral analysis effect may sum up to other effects (such as MSW
effect \cite{10} in the matter; in this connection we observe that the above
scheme is easily generalized to the oscillations in the matter, see ref.1)
in depressing or enhancing neutrino oscillations.

Finally, we remark that, as shown in ref. 1,
the ratio of the amplitudes of the
$|{\alpha_{k,1}}^{r}>$ and $|{\alpha_{k,2}}^{r}>$ components of the state
$|{\alpha_{k,e}}^{r}(t)>$ is constant in time and that such a feature persists
even in the relativistic limit, where, however, the oscillation formula
reduces to the usual one. This reminds us of the SU(2) coherent state structure
of the vacuum state \cite{1} and is in contrast with the conventional treatment
where the phase factor $exp{(-i\Delta \omega t)}$ produces
"decoherence" between the components $|{\alpha_{k,1}}^{r}>$ and
$|{\alpha_{k,2}}^{r}>$.

In conclusion, the above discussion shows that the momentum dependence of
the oscillation amplitude may be subject to experimental test and may
provide novel features with respect to the conventional treatment
of neutrino mixing. The vacuum condensate structure may manifest
itself through its phenomenological consequences on the neutrino
oscillations.

{\bf Acknowledgments}

We acknowledge useful discussions with S.M.Bilenky, R.Iengo and A.Perelomov.
This work has been partially supported by EU Contract ERB CHRX CT94 0423.


\newpage
\thispagestyle{empty}
\begin{center}

{\bf Figure caption}

\medskip

\bigskip

\bigskip

\bigskip

Fig. 1: The function $|U(p,a)|^{2}$.
\end{center}

\end{document}